\documentclass[12pt]{article}

\newcommand{\be}{\begin{equation}}
\newcommand{\ee}{\end{equation}}
\newcommand{\bi}[1]{\vspace{-3mm} \bibitem{#1}}

\usepackage{epsfig,amsmath,amssymb,graphics,graphicx} 

\begin{document}

\begin{center}
{\it Modern Physics Letters A 21 (2006) 1587-1600}
\end{center}

\begin{center}
{\Large \bf Electromagnetic Fields on Fractals}

\vskip 5 mm

{\large \bf Vasily E. Tarasov }

{\it Skobeltsyn Institute of Nuclear Physics, \\ 
Moscow State University, Moscow 119991, Russia } \\
{E-mail: tarasov@theory.sinp.msu.ru}
\end{center}

\begin{abstract}
Fractals are measurable metric sets with non-integer Hausdorff dimensions. 
If electric and magnetic fields are defined on fractal and 
do not exist outside of fractal in Euclidean space, 
then we can use the fractional generalization of the
integral Maxwell equations.
The fractional integrals are considered as approximations of
integrals on fractals. 
We prove that fractal can be described as a specific medium.
\end{abstract}




\section{Introduction}

The theory of integrals and derivatives of non-integer order goes back 
to Leibniz, Liouville, Riemann, Grunwald, and Letnikov \cite{SKM,OS}. 
Fractional analysis has found many
applications in recent studies in mechanics and physics.
The interest in fractional integrals and derivatives 
has been growing continuously 
during the last few years because of numerous applications. 
In a short period of time the list of applications has been expanding. 
It includes chaotic dynamics \cite{Zaslavsky1,Zaslavsky2},
physics of fractal and complex media 
\cite{Mainardi,Nig4,Hilfer,Media,Physica2005},
physical kinetics \cite{Zaslavsky1,Zaslavsky7,SZ,ZE},
plasma physics \cite{Lutzen,Mil2},  
astrophysics \cite{CMDA},
long-range dissipation \cite{GM,TZ2}, 
non-Hamiltonian mechanics \cite{nonHam,FracHam},
theory of long-range interactions \cite{Lask,TZ3,KZT}.

The new type of problem has increased rapidly in areas 
in which the fractal features of the process or the medium 
impose the necessity of using non-traditional tools. 
In order to use fractional derivatives and fractional integrals 
for media on fractal, we must use some 
continuous model \cite{Media}.  
We propose to describe the medium on fractal by a fractional 
continuous model \cite{Media}, where all characteristics and 
fields are defined 
everywhere in the volume but they follow some generalized 
equations, which are derived by using fractional integrals.
In many problems the real fractal structure 
can be disregarded and the medium on fractal can be described by
some fractional continuous mathematical model.
The order of the fractional integral is equal
to the fractal dimension. 
Fractional integrals can be considered as approximations 
of integrals on fractals \cite{Svozil,RLWQ}. 
In Ref. \cite{RLWQ}, authors proved that integrals
on a net of fractals can be approximated by fractional integrals.
In Ref. \cite{nonHam}, we proved that fractional integrals
can be considered as integrals over the space with non-integer
dimension up to numerical factor. 
This interpretation follows from the well-known formulas 
for dimensional regularization \cite{Col}.

In Sec. 2, a brief review of 
Hausdorff measure, Hausdorff dimension 
and integration on fractals suggested
to fix notation and provide a convenient reference.
The connection integration on fractals and fractional 
integration is discussed.
In Sec. 3, the fractional electrodynamics on 
fractals is considered. 
Fractional generalization of the
integral Maxwell equations is suggested. 
Finally, a short conclusion is given in Sec. 4.

\section{Integration on fractal and fractional integration} 

\subsection{Hausdorff measure and Hausdorff dimension}

Fractals are measurable metric sets with fractal Hausdorff dimension.  
The main property of fractal is non-integer Hausdorff dimension.
Let us consider a brief review of Hausdorff measure and 
Hausdorff dimension to fix notation and provide a convenient reference. 

Consider a measurable metric set $(W, \mu_H)$ with $W \subset \mathbb{R}^n$. 
The elements of $W$ are denoted by $x, y, z, . . . $, and represented by 
$n$-tuples of real numbers $x = (x_1,x_2,...,x_n)$ 
such that $W$ is embedded in $\mathbb{R}^n$. 
The set $W$ is restricted by the conditions:
(1) $W$ is closed;
(2) $W$ is unbounded; 
(3) $W$ is regular (homogeneous, uniform) with its points randomly distributed.

The diameter of a subset $E \subset W \subset \mathbb{R}^n$ is 
\[
d(E)=diam(E)=sup\{ d(x,y): x , y  \in E \}  ,
\]
where $d(x,y)$ is a metric function of two points: $x$ and $y \in W$. 

Let us consider a set $\{E_i\}$ of subsets $E_i$ such that
$diam(E_i) < \varepsilon$ for all $i \in \mathbb{N}$, and 
$W \subset \bigcup^{\infty}_{i=1} E_i$.
Then, we define 
\be
\xi(E_i,D)= \omega(D) [diam (E_i)]^D=\omega(D)  [d(E_i)]^D
\ee
for non-empty subsets $E_i$ of $W$. The factor $\omega(D)$ 
depends on the geometry of $E_i$, used for covering $W$.
If $\{E_i\}$ is the set of all (closed or open) balls in $W$, then
\be
\omega(D)=\frac{\pi^{D/2} 2^{-D}}{\Gamma(D/2+1)}. 
\ee

The Hausdorff dimension $D$ of a subset $E \subset W$ 
is defined \cite{Federer,R,Ed,Falconer} by 
\be \label{Hd}
D= dim_H (E)=\sup \{ d \in \mathbb{R}: \  \mu_H (E, d ) = \infty  \} 
=\inf\{ d \in \mathbb{R} : \ \mu_H(E,d)=0 \}.
\ee
The Hausdorff measure $\mu_H$ of a subset $E \subset W $ 
is defined \cite{Federer,R,Ed,Falconer} by
\be 
\mu_H(E,D)= \omega(D) \lim_{d(E_i) \rightarrow 0+} 
\inf_{\{E_i\}} \sum^{\infty}_{i=1} [d(E_i)]^D .
\ee
Note that $\mu_H(\lambda E,D) =\lambda^D \mu_H(E,D)$, 
where $\lambda >0$, and $\lambda E =\{ \lambda x, \ x \in E \}$.

\subsection{Function and integrals on fractal}

Let us consider the functions on $W$: 
\be \label{f}
f(x)=\sum^{\infty}_{i=1} \beta_i \chi_{E_i}(x) , 
\ee
where $\chi_{E}$ is the characteristic function of $E$: 
$$ \chi_{E}(x)=
\begin{cases} 
1 &  if \quad x \in E, \cr 
0 & if \quad x \not \in E
\end{cases}
$$

The Lebesgue-Stieltjes integral for (\ref{f}) is defined by
\be \label{LSI}
\int_W f d \mu =\sum^{\infty}_{i=1} \beta_i \mu_H(E_i).
\ee
Therefore
\be \label{int}
\int_W f(x) d \mu_H (x) =
\lim_{ d(E_i) \rightarrow 0} \sum_{E_i} f(x_i) \xi(E_i,D)= 
\omega(D) \lim_{ d(E_i) \rightarrow 0} \sum_{E_i} f(x_i) [d(E_i)]^D .
\ee
It is possible to divide $\mathbb{R}^n$ into parallelepipeds 
\be
E_{i_1...i_n} =\{  (x_1,...,x_n) \in W: \quad x_j =
(i_j-1) \Delta x_j +\alpha_j, \quad
0 \le \alpha_j \le \Delta x_j, \quad j=1,...,n \} .
\ee
Then
\be
d \mu_H (x)= \lim_{d(E_{i_1...i_n}) \rightarrow 0}
\xi(E_{i_1 ... i_n},D)=
\lim_{d(E_{i_1 ... i_n}) \rightarrow 0}
\prod^n_{j=1} (\Delta x_j)^{D/n}=\prod^n_{j=1} d^{D/n} x_j .
\ee
The range of integration $W$ can be parametrized by 
polar coordinates with $r=d(x, 0)$ and angle $\Omega$. 
Then $E_{r,\Omega}$ can be thought of as spherically 
symmetric covering around a center at the origin. 
In the limit, the function $\xi(E_{r,\Omega},D)$ gives
\be
d\mu_H(r,\Omega)=\lim_{d(E_{r, \Omega}) \rightarrow 0}
\xi (E_{r,\Omega},D)=d\Omega^{D-1} r^{D-1} dr. 
\ee

Let us consider $f(x)$ that is symmetric with respect 
to some center $x_0 \in W$, 
i.e. $f(x) = const$ for all $x$ such that $d(x, x_0)=r$ 
for arbitrary values of $r$. Then the transformation
\be \label{WrZ}
W \rightarrow W^{\prime} : \quad x \rightarrow x^{\prime}=x-x_0
\ee
can be performed to shift the center of symmetry.  
Since $W$ is not a linear space, 
(\ref{WrZ}) need not be a map of $W$ onto itself. 
The map (\ref{WrZ}) is measure-preserving. 
Then the  integral over a $D$-dimensional metric space is 
\be \label{intWf}
\int_W f d\mu_H = \frac{2 \pi^{D/2}}{\Gamma(D/2)}  
\int^{\infty}_0 f(r) r^{D-1} dr .
\ee
This integral is known in the theory of the 
fractional calculus \cite{SKM}. 
The right Riemann-Liouville fractional integral is
\be \label{FID}
I^{D}_{-} f(z)=\frac{1}{\Gamma(D)} \int^{\infty}_z (x-z)^{D-1} f(x) dx .
\ee
Equation (\ref{intWf}) is reproduced by
\be \label{FIFI}
\int_W f d\mu_H = \frac{2 \pi^{D/2} \Gamma(D)}{\Gamma(D/2)} I^{D}_{-} f(0) .
\ee
Relation (\ref{FIFI}) connects the integral on fractal 
with integral of fractional order.
This result permits to apply different tools of the fractional calculus
\cite{SKM} for the fractal medium.
As a result, the fractional integral can be considered as an
integral on fractal (fractional Hausdorff dimension set)
up to the numerical factor $\Gamma(D/2) /[ 2 \pi^{D/2} \Gamma(D)]$.

Note that the interpretation of fractional integration
is connected with fractional dimension \cite{nonHam}.
This interpretation follows from the well-known formulas 
for dimensional regularization \cite{Col,Wilson}.
The fractional integral can be considered as an
integral in the fractional dimension space
up to the numerical factor $\Gamma(D/2) /[ 2 \pi^{D/2} \Gamma(D)]$.
In Ref. \cite{Svozil} was proved that the fractal space-time 
approach is technically identical with dimensional regularization.

The integral defined in (\ref{int}) satisfies the 
translational invariance property:
\be \label{15}
\int_W f(x+x_0) d \mu_H(x)= \int_W f(x) d \mu_H(x)
\ee
since $d \mu_H(x - x_0)=d \mu_H(x)$ as a consequence of homogeneity.
The integral (\ref{int}) satisfies the scaling property:
\be \label{16}
\int_W f(\lambda x) d \mu_H(x)= \lambda^{-D}\int_W f(x) d \mu_H(x)
\ee
since $d \mu_H (x/\lambda)=\lambda^{-D}d \mu_H(x)$.

\subsection{Multi-variable integration on fractal}

The integral in (\ref{intWf}) is defined for a single variable 
but not multiple variables. 
It is only useful for integrating spherically symmetric functions. 
We consider multiple 
variables by using the product spaces and product measures.

Let us consider a collection of $n=3$ 
measure spaces $(W_k,\mu_k,D)$ with $k=1,2,3$,
and form a Cartesian product of the sets $W_k$ producing the space 
$W=W_1 \times W_2 \times W_3$.
The definition of product measures and application of Fubini's theorem 
provides a measure for the product set $W=W_1 \times W_2 \times W_3$ as
\be
(\mu_1 \times \mu_2 \times \mu_3)(W) 
= \mu_1(W_1) \mu_2(W_2) \mu_3(W_3). 
\ee
The integration over a function $f$ on the product space is
\be \label{int-n}
\int f ({\bf r}) d(\mu_1 \times \mu_2 \times \mu_3) = 
\int \int \int f (x_1,x_2 , x_3) d \mu_1(x_1) d \mu_2(x_2) d \mu_3(x_3). 
\ee
In this form, the single-variable measure from (\ref{intWf}) may 
be used for each coordinate $x_k$, which has 
an associated dimension $\alpha_k$:
\be 
d \mu_k(x_k) = \frac{2 \pi^{\alpha_k/2}}{\Gamma(\alpha_k/2)} 
|x_k|^{\alpha_k-1} dx_k , \quad k=1,2,3.
\ee
The total dimension of  $W=W_1 \times W_2 \times W_3$ is  
$D= \alpha_1+\alpha_2+\alpha_3$. 

Let us reproduce the result for the single-variable integration 
(\ref{intWf}), from the product space $W_1  \times  W_2 \times W_3$.  
We take a spherically symmetric function 
$f({\bf r})=f (x_1, x_2 , x_3) = f (r)$,
where $r^2 = (x_1)^2 + (x_2)^2 + (x_3)^2$, and
to perform the integration in spherical coordinates $(r, \phi, \theta)$,
we use
\be
\int^{\pi/2}_0 \sin^{\mu-1} x \cos^{\nu-1} x dx =
\frac{\Gamma(\mu/2) \Gamma(\nu/2)}{2 \Gamma((\mu+\nu)/2)} ,
\ee
where $\mu >0$, $\nu>0$. Then Eq. (\ref{int-n}) becomes
\be
\int d\mu_1(x_1) d \mu_2(x_2) d\mu_3(x_3)f (r) = 
\frac{2 \pi^{D/2}}{\Gamma(D/2)} \int f(r) r^{D-1} dr .
\ee
This equation describes integration over 
a spherically symmetric function 
in a $D$-dimensional space and 
reproduces the result (\ref{intWf}).

\section{Electrodynamics on fractal}

\subsection{Electric charge of fractal set}

Let us consider the electric charge that 
is distributed on the measurable metric set 
$W$ with the fractional Hausdorff dimension $D$.
Suppose that the density of charge distribution is described 
by the function $\rho({\bf r},t)$.
In this case, the total charge is defined by 
\be \label{QD}
Q_D(W)=\int_W \rho ({\bf r},t) dV_D, \quad 
dV_D=d \mu_1(x_1) d \mu_2(x_2) d \mu_3(x_3)=c_3(D,{\bf r}) dV_3,
\ee 
where $dV_3=dx dy dz$ for Cartesian coordinates, 
$dim_H (W)=D =\alpha_1+\alpha_2+\alpha_3$, and
\be
c_3(D,{\bf r})=\frac{8 \pi^{D/2}
|x|^{\alpha_1-1} |y|^{\alpha_2-1} |z|^{\alpha_3-1} }{\Gamma(\alpha_1) 
\Gamma(\alpha_2) \Gamma(\alpha_3) } .
\ee
As a result, we get Riemann-Liouville fractional integral \cite{SKM}
up to numerical factor $8 \pi^{D/2}$.
Note that the final equations that relate the physical variables 
have the form that 
are independent of numerical factor in the function $c_3(D,{\bf r})$. 
However, the dependence of ${\bf r}$ is important to these equations. 

Equation (\ref{QD}) describes the charge that is distributed in the volume
and has the fractal dimension $D$ by fractional integrals.
There are many different definitions of fractional integrals \cite{SKM}.
For the Riemann-Liouville fractional integral, 
the function $c_3(D,{\bf r})$ is 
\be \label{c3Dr}
c_3(D,{\bf r})=\frac{ |x|^{\alpha_1-1}|y|^{\alpha_2-1} |z|^{\alpha_3-1} }{
\Gamma(\alpha_1) \Gamma(\alpha_2) \Gamma(\alpha_3) },
\ee
where $x$, $y$, $z$ are Cartesian's coordinates, 
and $D=\alpha_1+\alpha_2+\alpha_3$, $0<D\le 3$.
Note that for $D=2$, we have the distribution in the volume. 
In general, this case is not equivalent to the distribution
on the two-dimensional surface.
For $\rho({\bf r})=\rho(|{\bf r}|)$, we can use
the fractional integrals with
\be \label{IDc}
c_3(D,{\bf r})=\frac{2^{3-D} \Gamma(3/2)}{\Gamma(D/2)} |{\bf r}|^{D-3}  .
\ee

If we consider the ball region $W=\{{\bf r}: \ |{\bf r}|\le R \}$, 
and stationary spherically symmetric distribution of charged particles 
($\rho({\bf r},t)=\rho(r)$), then 
\[ Q_D(R)=4\pi \frac{2^{3-D}\Gamma(3/2)}{\Gamma(D/2)}
\int^R_0 \rho(r) r^{D-1} dr . \]
For the homogeneous case, $\rho(r,t)=\rho_0$, and 
\[ Q_D(R)=4\pi \rho_0 \frac{2^{3-D}\Gamma(3/2)}{\Gamma(D/2)}
\frac{R^D}{D} \sim R^D . \]
The distribution of charged particles is homogeneous 
if all regions $W$ and $W^{\prime}$ with equal 
volumes $V_D(W)=V_D(W^{\prime})$ have the same 
total charges on these regions  $Q_D(W)=Q_D(W^{\prime})$. 

For charged particles that are distributed with a constant density 
over a fractal with Hausdorff dimension $D$, 
the electric charge $Q$ satisfies the scaling law $Q(R) \sim R^{D}$,
whereas for a regular n-dimensional Euclidean object 
we have $Q(R)\sim \mathbb{R}^n$.

\subsection{Electric current for fractal}

For charged particles with density $\rho({\bf r},t)$ flowing 
with velocity ${\bf u}={\bf u}({\bf r},t)$, 
the resulting current density ${\bf J}({\bf r},t)$ is 
\[ {\bf J}({\bf r},t)= \rho({\bf r},t) {\bf u}({\bf r},t) . \]
The electric current $I(S)$ is defined as the flux of electric charge.
Measuring the field ${\bf J}({\bf r},t)$ passing through a surface 
$S=\partial W$ gives  
\be \label{IS0}
I(S)=\Phi_J(S)=\int_S ({\bf J} ({\bf r},t), d{\bf S}_2) , \ee
where $d{\bf S_2}=dS_2{\bf n}$ is a differential unit of area 
pointing perpendicular to the surface $S$, 
and the vector ${\bf n}=n_k {\bf e}_k$ is a vector of normal.
The fractional generalization of (\ref{IS0}) is
\be \label{IS}
I(S)=\int_S ({\bf J}({\bf r},t), d{\bf S}_d) , 
\ee
where 
\be \label{C2} d{\bf S}_d=c_2 (d,{\bf r})d{\bf S}_2 , \quad 
c_2(d,{\bf r})= \frac{2^{2-d}}{\Gamma(d/2)} |{\bf r}|^{d-2} . \ee
Note that $c_2(2,{\bf r})=1$ for $d=2$. 
The boundary $\partial W$ has the dimension $d$. 
In general, the dimension $d$ is not equal to $2$ and 
is not equal to $(D-1)$.

\subsection{Charge conservation for fractal}

The electric charge has a fundamental property established
by numerous experiments: the velocity of charge change
in region $W$ bounded by the surface $S=\partial W$
is equal to the flux of charge through this surface.
This is known as the law  of charge conservation:
\[ \frac{dQ(W)}{dt}=-I(S), \]
or, in the form
\be \label{cecl} \frac{d}{dt} \int_W \rho({\bf r},t) dV_D= 
- \oint_{\partial W} ({\bf J} ({\bf r},t),d{\bf S}_d) . \ee
In particular, when the surface $S=\partial W$ is fixed,
we can write
\be \label{drho} \frac{d}{dt} \int_W \rho({\bf r},t) dV_D= 
\int_W \frac{\partial \rho({\bf r},t)}{\partial t} dV_D .\ee
Using the fractional generalization of the 
Gauss's theorem (see the Appendix), we get
\be \label{gt}
\oint_{\partial W} ({\bf J} ({\bf r},t),d{\bf S}_d) 
=\int_W c^{-1}_3(D,{\bf r})
\frac{\partial}{\partial x_k} \Bigl( c_2(d,{\bf r})J_k({\bf r},t) \Bigr)
dV_D .\ee
Substitution of Eqs. (\ref{drho}) and (\ref{gt})
into Eq. (\ref{cecl}) gives 
\be \label{53}
c_3(D,{\bf r})\frac{\partial \rho({\bf r},t)}{\partial t}+
\frac{\partial}{\partial x_k} \Bigl( c_2(d,{\bf r})J_k({\bf r},t) \Bigr)=0. 
\ee
As a result, we obtain the law of charge 
conservation in differential form (\ref{53}).
This equation can be considered as a continuity equation for
fractal.

\subsection{Electric field and Coulomb's law}

For a continuous stationary distribution $\rho({\bf r}^{\prime})$,   
the electric field at a point ${\bf r}$ is defined by
\be \label{E}
{\bf E}({\bf r})=\frac{1}{4 \pi \varepsilon_0} \int_W
\frac{{\bf r}-{\bf r}^{\prime}}{|{\bf r}-{\bf r}^{\prime}|^3}
\rho({\bf r}^{\prime}) dV^{\prime}_3 ,
\ee
where $\varepsilon_0$
is a fundamental constant called the permittivity of free space. 
For Cartesian's coordinates 
$dV^{\prime}_3=dx^{\prime}dy^{\prime}dz^{\prime}$.
The fractional generalization of (\ref{E}) is
\be \label{CLD}
{\bf E}({\bf r})=\frac{1}{4 \pi \varepsilon_0} \int_W
\frac{{\bf r}-{\bf r}^{\prime}}{|{\bf r}-{\bf r}^{\prime}|^3}
\rho({\bf r}^{\prime}) dV^{\prime}_D , \ee
where $dV^{\prime}_D=c_3(D,{\bf r}^{\prime}) dV^{\prime}_3$. 
Equation (\ref{CLD}) can be considered as Coulomb's law 
for a stationary distribution of electric charges on fractal.

\subsection{Gauss's law for fractal}

The Gauss's law tells us that the total flux $\Phi_E(S)$ of 
the electric field ${\bf E}$
through a closed surface $S=\partial W$ 
is proportional to the total electric charge $Q(W)$
inside the surface: 
\be \label{GL1} \Phi_E(\partial W)=\frac{1}{\varepsilon_0} Q(W) . \ee
The electric flux for the surface $S=\partial W$ is
\[ \Phi_E(S)=\int_S ({\bf E}, d{\bf S}_2) , \]
where ${\bf E}({\bf r},t)$ is the electric field vector, and $d{\bf S}_2$ 
is a differential unit of area pointing perpendicular to the surface $S$. 

For the distribution on fractal, the Gauss's law (\ref{GL1}) states 
\be \label{GL2} \int_S ({\bf E},d{\bf S}_2)=\frac{1}{\varepsilon_0} 
\int_W \rho ({\bf r},t) dV_D , \ee
where $\rho({\bf r},t)$ is the 
density of electric charge that is distributed on fractal,  
$dV_D=c_3(D,{\bf r})dV_3$,
and $\varepsilon_0$ is the permittivity of free space.

If $\rho({\bf r},t)=\rho(r)$, and $W=\{{\bf r}:\ |{\bf r}|\le R\}$, then 
\be \label{QW2}
Q(W)=4 \pi \int^R_0 \rho(r) c_3(D,{\bf r}) r^2 dr =
4 \pi \frac{2^{3-D}\Gamma(3/2)}{\Gamma(D/2)}
\int^R_0 \rho(r) r^{D-1} dr . \ee
For the sphere $S=\partial W=\{{\bf r}: \ |{\bf r}|= R \}$,
\be \label{PW} \Phi_E(\partial W)= 4 \pi R^2 E(R). \ee
Substituting (\ref{QW2}) and (\ref{PW}) in (\ref{GL1}),
we get 
\[ E(R)=\frac{2^{3-D}\Gamma(3/2)}{\varepsilon_0 R^2 \Gamma(D/2)}
\int^R_0 \rho(r) r^{D-1} dr .\]
For homogeneous ($\rho(\bf r)=\rho$) distribution, 
\[ E(R)=\rho \frac{2^{3-D}\Gamma(3/2)}{\varepsilon_0 D \Gamma(D/2)} 
R^{D-2}  \sim R^{D-2} .\]

\subsection{Magnetic field and Biot-Savart law}
    
The Biot-Savart law relates magnetic fields to the currents, 
which are their sources. 
For a continuous distribution, the law is
\be \label{BSL0} {\bf B}({\bf r})=\frac{\mu_0}{4\pi} \int_W 
\frac{[{\bf J}({\bf r}^{\prime}),{\bf r}-{\bf r}^{\prime}]}{
|{\bf r}-{\bf r}^{\prime}|^3} d V^{\prime}_3 , \ee
where $[\ , \ ]$ is a vector product, 
${\bf J}$ is the current density, 
$\mu_0$ is the permeability of free space. 
The fractional generalization of Eq. (\ref{BSL0}) is
\be \label{BSL} {\bf B}({\bf r})=\frac{\mu_0}{4\pi} \int_W 
\frac{[{\bf J}({\bf r}^{\prime}),{\bf r}-{\bf r}^{\prime}]}{
|{\bf r}-{\bf r}^{\prime}|^3} d V^{\prime}_D . \ee
This equation is Biot-Savart law written 
for a steady current with fractal distribution of electric charges.
The law (\ref{BSL}) can be used to find the magnetic 
field produced by any distribution of steady currents on fractal.

\subsection{Ampere's law for fractal}

The magnetic field in space around an electric current 
is proportional to the electric current, which serves as its source. 
In the case of static electric field, 
the line integral of the magnetic field around 
a closed loop is proportional to the electric current 
flowing through the loop. 
The Ampere's law 
is equivalent to the steady state of the integral Maxwell equation 
in free space, and relates the spatially varying magnetic field 
${\bf B}({\bf r})$ to the current density ${\bf J}({\bf r})$. 

Note that, as mentioned in Ref. \cite{Lutzen}, 
Liouville, who was one of the pioneers in developing  
fractional  calculus, was inspired by the problem of 
fundamental force law in Ampere's electrodynamics and 
used fractional differential equation in that  problem.  

The Ampere's law states that
the line integral of the magnetic field ${\bf B}$ along 
the closed path $L$ around a current given in MKS by
\[ \oint_L ({\bf B},d{\bf l})=\mu_0 I(S) , \]
where $d{\bf l}$ is the differential length element.  
For the distribution of particles on the fractal, 
$I(S)$ is defined by (\ref{IS}). 
For the cylindrically symmetric distribution, 
\[ I(S)=2 \pi \int^R_0 J(r) c_2(d,{\bf r}) r dr 
=4 \pi \frac{2^{2-d}}{\Gamma(d/2)}
\int^R_0 J(r) r^{d-1} dr , \]
where we use $c_2(d,{\bf r})$ from Eq. (\ref{C2}).  
For the circle $L=\partial W=\{{\bf r}: \ |{\bf r}|=R \}$, we get
\[ \oint_L ({\bf B},d{\bf l})= 2 \pi R \ B(R). \]
As a result, 
\[ B(R)= \frac{ \mu_0 2^{2-d}}{R \Gamma(d/2)}
\int^R_0 J(r) r^{d-1} dr .\]
For the homogeneous case, $J(r)=J_0$ and
\[ B(R)=J_0 \frac{\mu_0 2^{2-d}}{d \Gamma(d/2)} R^{d-1} \sim R^{d-1} .\]

\subsection{Fractional integral Maxwell equations} 

Let us consider the fractional integral Maxwell equations.
The Maxwell equations are the set of 
fundamental equations for electric and magnetic fields.
The equations can be expressed in integral form 
known as Gauss's law, Faraday's law, 
the absence of magnetic monopoles, and Ampere's law 
with displacement current.
In MKS,  these become 
\[ \oint_S ({\bf E},d{\bf S}_2)=
\frac{1}{\varepsilon_0} \int_W \rho dV_D , 
\]
\[ \oint_L ({\bf E},d{\bf l}_1)=
-\frac{\partial}{\partial t} \int_S ({\bf B},d{\bf S}_2) , \]
\[ \oint_S ({\bf B},d{\bf S}_2)= 0, \]
\[
\oint_L ({\bf B},d{\bf l}_1)=\mu_0 \int_S ({\bf J}, d{\bf S}_d)
+ \varepsilon_0 \mu_0\frac{\partial}{\partial t}
\int_S ({\bf E},d{\bf S}_2) . \]
Let us consider the fields that are defined on fractal only.
The hydrodynamic and thermodynamics fields 
can be defined in the fractal media \cite{Media,Physica2005}.
Suppose that the electromagnetic field can be defined on fractal 
as an approximation of some real case with fractal medium.
If the electric field ${\bf E}({\bf r},t)$ and magnetic 
fields ${\bf B}({\bf r},t)$ can be defined on fractal and 
do not exist outside of fractal in Euclidean space $\mathbb{R}^3$, 
then we must use the fractional generalization of the
integral Maxwell equations in the form:
\be \label{41}
\oint_S ({\bf E},d{\bf S}_d)=
\frac{1}{\varepsilon_0} \int_W \rho dV_D ,
\ee
\be 
\oint_L ({\bf E},d{\bf l}_{\gamma})=
-\frac{\partial}{\partial t} \int_S ({\bf B},d{\bf S}_d) , 
\ee
\be
\oint_S ({\bf B},d{\bf S}_d)= 0, 
\ee
\be \label{Max1}
\oint_L ({\bf B},d{\bf l_{\gamma}})=\mu_0 \int_S ({\bf J}, d{\bf S}_d)
+ \varepsilon_0 \mu_0\frac{\partial}{\partial t}
\int_S ({\bf E},d{\bf S}_d) . \ee
The fractional integrals are considered as approximation 
of integrals on fractals \cite{Svozil,RLWQ}. 

Using the fractional generalization of Stokes's and Gauss's
theorems (see the Appendix), 
we can rewrite Eqs. (\ref{41}) - (\ref{Max1}) in the form
\be
\int_W c^{-1}_3(D,{\bf r}) div( c_2(d,{\bf r}) {\bf E}) dV_D 
= \frac{1}{\varepsilon_0} \int_W \rho dV_D , \ee
\be \int_S  c^{-1}_2(d,{\bf r})
( curl(c_1(\gamma,{\bf r}){\bf E}), d{\bf S}_d) = 
-\frac{\partial}{\partial t} \int_S ({\bf B},d{\bf S}_d) , 
\ee
\be 
\int_W c^{-1}_3(D,{\bf r}) div(c_2(d,{\bf r}) {\bf B}) dV_d=0, 
\ee
\be 
\int_S c^{-1}_2(d,{\bf r}) 
(curl(c_1(\gamma,{\bf r}){\bf B}), d{\bf S}_d) = 
\mu_0 \int_S ({\bf J}, d{\bf S}_d) 
+ \varepsilon_0 \mu_0\frac{\partial}{\partial t}
\int_S ({\bf E},d{\bf S}_d) .  
\ee
As a result, we obtain  
\be \label{M1}
div \Bigl(c_2(d,{\bf r}) {\bf E} \Bigr) = 
\frac{1}{\varepsilon_0} c_3(D,{\bf r}) \rho  , 
\ee
\be \label{M2}
curl \Bigl(c_1(\gamma,{\bf r}){\bf E} \Bigr)= 
-c_2(d,{\bf r}) \frac{\partial}{\partial t} {\bf B} , 
\ee
\be \label{M3}
div \Bigl( c_2(d,{\bf r}) {\bf B} \Bigr)= 0, 
\ee
\be  \label{M4}
curl \Bigl( c_1(\gamma,{\bf r}){\bf B} \Bigr) = 
\mu_0 c_2(d,{\bf r}) {\bf J}+\varepsilon_0 \mu_0 c_2(d,{\bf r})
\frac{\partial {\bf E}}{\partial t}. \ee

Note that the law of absence of magnetic monopoles 
for the fractal leads us to the equation 
$div ( c_2(d,{\bf r}) {\bf B} )= 0 $. 
It can be rewritten as
\[ div{\bf B}=-({\bf B}, grad c_2(d,{\bf r}) ) . \]
In the general case ($d \not=2$), the vector
$grad \ (c_2(d,{\bf r}))$ is not equal 
to zero and the magnetic field satisfies $div {\bf B}\not=0$.
If $d=2$, we have $div ({\bf B})\not=0$ only for non-solenoidal 
field ${\bf B}$. Therefore the magnetic field on the fractal  
is similar to the non-solenoidal field. 
As a result, the field on fractal can be considered 
as a field with some "fractional magnetic monopole" \ 
$q_m\sim ({\bf B},\nabla c_2)$.

\subsection{Fractal as effective medium}

The Maxwell equations (\ref{M1})-(\ref{M4}) on fractal
can be considered as the equations for medium
\be \label{M1b}
div \Bigl( {\bf D} \Bigr) = \rho^{eff}_{free} , 
\ee
\be \label{M2b}
curl \Bigl({\bf E}^{eff} \Bigr)= 
- \frac{\partial}{\partial t} {\bf B}^{eff} , 
\ee
\be \label{M3b}
div \Bigl( {\bf B}^{eff} \Bigr)= 0, 
\ee
\be  \label{M4b}
curl \Bigl( {\bf H} \Bigr) = 
{\bf J}^{eff}+\frac{\partial {\bf D}}{\partial t}. \ee

The effective Maxwell equations (\ref{M1b})-(\ref{M4b})
prove that fractal creates some polarization and magnetization.
In the equations, we use some effective fields
\be
{\bf E}^{eff}({\bf r},t)=c_1(\gamma,{\bf r}){\bf E}({\bf r},t), \quad
{\bf B}^{eff}({\bf r},t)=c_2(d,{\bf r}) {\bf B}({\bf r},t) .
\ee
The fields ${\bf E}^{eff}$ and ${\bf B}^{eff}$ 
mean that electromagnetic fields ${\bf E}$ and ${\bf B}$ 
of particles are changed by fractal.

Equations (\ref{M1b})-(\ref{M4b}) have
the effective free charge and current densities:
\be
\rho^{eff}_{free}({\bf r},t)=c_3(D,{\bf r}) \rho ({\bf r},t), \quad 
{\bf J}^{eff}({\bf r},t)=c_2(d,{\bf r}) {\bf J}({\bf r},t) .
\ee
We can interpret the existence of $\rho^{eff}_{free}$ and ${\bf J}^{eff}$
in the equations as an effect of change of the free 
electric charge and current densities by fractal. 
This change exists in addition to the effect of appearance
the dipole charges and polarization or magnetization currents.
The fractal can be considered as a medium that has
the electrical and magnetic permittivities in the form
\be
\varepsilon=c_2(d,{\bf r}) c^{-1}_1(\gamma,{\bf r}), \quad
\mu=c_2(d,{\bf r}) c^{-1}_1(\gamma,{\bf r}).
\ee
The fields ${\bf D}$ and ${\bf H}$ are related to ${\bf E}^{eff}$ 
and ${\bf B}^{eff}$ by the usual equations:
\be
{\bf D}=\varepsilon \varepsilon_0 {\bf E}^{eff},
\quad
{\bf H}=\frac{1}{\mu \mu_0} {\bf B}^{eff} .
\ee
Note that the continuity equation (\ref{53}) for
fractal can be presented by
\be 
\frac{\partial \rho^{eff}({\bf r},t)}{\partial t}+
div \Bigl( {\bf J}^{eff}({\bf r},t) \Bigr)=0. 
\ee

As a result, the fractal can be considered as a specific
medium that changes the fields, free charges and currents
in addition to the creation of polarization and magnetization.


\section{Conclusion}

Fractals are measurable metric sets with non-integer 
Hausdorff dimensions. We consider the electric and magnetic 
fields that are defined on fractal and 
does not exist outside of fractal in Euclidean space. 
For charged particles that are distributed with a constant 
density over a fractal with Hausdorff dimension $D$, 
the electric charge $Q$ satisfies the scaling law $Q(R) \sim R^{D}$,
whereas for a regular n-dimensional Euclidean object 
we have $Q(R)\sim \mathbb{R}^n$. 
This property can be used to measure the 
fractal Hausdorff dimension $D$.

The fractional integrals can be used to
describe electromagnetic fields on fractals.
These integrals are considered as approximations of
integrals on fractals. 
The fractional generalizations of integral Maxwell equations
for fractal set are derived.
The magnetic field on fractal can be considered as 
a field with some "fractional magnetic monopole". 

We can interpret the equations for electromagnetic fields on fractal
as an effect of creation  of some polarization and magnetization
by fractal. 
Moreover, the electromagnetic fields  are also changed by fractal.
From the generalized Maxwell equations, we can see 
the effect of change of the free 
electric charge and current densities by fractal. 
This change exists in addition to the effect of appearance
the dipole charges and polarization or magnetization currents.
The electrical permittivity $\varepsilon$ 
and the magnetic permittivity $\mu$ of fractal
are defined by the Hausdorff measure and dimension of fractal.


\section*{Appendix: Fractional Gauss's theorem}

Let us derive the fractional 
generalization of the Gauss's theorem
\be \label{AAA} 
\int_{\partial W} ({\bf J}({\bf r},t), d{\bf S}_2) 
=\int_W div( {\bf J}({\bf r},t) ) dV_3 , 
\ee
where the vector ${\bf J}({\bf r},t)=J_k{\bf e}_k$ is a field, 
and 
\[ div( {\bf J})={\partial {\bf J}}/{\partial {\bf r}}= 
{\partial J_k}/{\partial x_k} . \]
Here, we mean the sum on the repeated index
$k$ from 1 to 3. Using 
\[ d{\bf S}_d=c_2 (d,{\bf r})d{\bf S}_2 , \quad 
c_2(d,{\bf r})= \frac{2^{2-d}}{\Gamma(d/2)} |{\bf r}|^{d-2} , \]
we get
\[ \int_{\partial W} ({\bf J}({\bf r},t),d{\bf S}_d) 
=\int_{\partial W}  c_2(d,{\bf r})  ({\bf J}({\bf r},t) , d{\bf S}_2) . \]
Note that $c_2(2,{\bf r})=1$ for $d=2$. 
Using (\ref{AAA}), we get 
\[ \int_{\partial W}  c_2(d,{\bf r}) ({\bf J}({\bf r},t), d{\bf S}_2) =
\int_W  div(c_2(d,{\bf r}) {\bf J}({\bf r},t)) dV_3 . \]
The relation $dV_3=c^{-1}_3(D,{\bf r}) dV_D$
allows us to derive the fractional generalization of the Gauss's theorem:
\[ \int_{\partial W} ({\bf J}({\bf r},t), d{\bf S}_d)
=\int_W c^{-1}_3(D,{\bf r}) 
div \Bigr( c_2(d,{\bf r}) {\bf J}({\bf r},t) \Bigr) \ dV_D .\]
Analogously, we can get the fractional generalization
of Stokes's theorem in the form
\[ \oint_L ({\bf E},d{\bf l}_{\gamma})=
\int_S  c^{-1}_2(d,{\bf r})
(curl(c_1(\gamma,{\bf r}){\bf E}), d{\bf S}_d) , \]
where 
\[ c_1(\gamma,{\bf r})=
\frac{2^{1-\gamma}\Gamma(1/2)}{\Gamma(\gamma/2)}|{\bf r}|^{\gamma-1} . \]


\end{document}